\documentclass[conference,compsoc]{IEEEtran}

\ifCLASSOPTIONcompsoc
  \usepackage[nocompress]{cite}
\else
  \usepackage{cite}
\fi

\usepackage{graphicx}
\graphicspath{{figures/}}
\DeclareGraphicsExtensions{.jpg,.png}
\usepackage{etoolbox}
\usepackage[caption=false,font=footnotesize]{subfig}

\usepackage{multirow,makecell}

\usepackage{framed}
\usepackage{float}

\usepackage[nodayofweek]{datetime}
\usepackage{listings}

\usepackage{array}
\usepackage{rotating}
\usepackage{adjustbox}
\usepackage{tabularx}
\usepackage[framemethod=default]{mdframed}
\usepackage{tikz}
\usepackage{pgfplots, pgfplotstable}
\usetikzlibrary{patterns}
\usepackage{booktabs}

\usepackage{rotating}
\usepackage{framed}
\usepackage{float}
\newfloat{Protocol}{t}{struct}

\usepackage{makecell}
\usepackage{multirow}
\usepackage{comment}

\usepackage[ruled,vlined]{algorithm2e}
\usepackage{algpseudocode}
\usepackage{tikz}
\usepackage{pgfplots}
\usetikzlibrary{patterns}

\usepackage{amssymb}
\usepackage{listings}
\usepackage{pifont}
\makeatletter
\def\blfootnote{\xdef\@thefnmark{}\@footnotetext}
\makeatother

\usepackage{pgfplotstable}
\usepackage{filecontents}

\usepackage{array}
\usepackage{xcolor,colortbl}
\usepackage[hidelinks,pdftitle={Title},pdfauthor={anonymous},unicode]{hyperref}
\usepackage{adjustbox}
\usepackage{fancyvrb}
\usepackage{balance}
\usepackage{blindtext}

\usepackage{xspace}
\newcommand{\etal}{\textit{et al.}\xspace}
\newcommand{\etc}{\textit{etc}\xspace}
\newcommand{\ie}{\textit{i.e.,}\xspace}
\newcommand{\eg}{\textit{e.g.,}\xspace}
\newcommand{\cf}{\textit{cf.}\xspace}

\newcounter{protocol}[section]

\newcolumntype{B}[2]{    >{\adjustbox{angle=#1,lap=\width-(#2)}\bgroup}
l    <{\egroup}}
\newcommand*\rotl{\multicolumn{1}{B{35}{1em}}}

\newcolumntype{Z}[2]{    >{\adjustbox{angle=#1,lap=\width-(#2)}\bgroup}
l    <{\egroup}}
\newcommand*\rott{\multicolumn{1}{Z{90}{1em}}}
\newcommand*\rots{\multicolumn{1}{Z{90}{1em}|}}

\newcolumntype{L}[1]{>{\raggedright\let\newline\\\arraybackslash\hspace{0pt}}m{#1}}
\newcolumntype{C}[1]{>{\centering\let\newline\\\arraybackslash\hspace{0pt}}m{#1}}
\newcolumntype{R}[1]{>{\raggedleft\let\newline\\\arraybackslash\hspace{0pt}}m{#1}}

\newcommand{\full}{$\bullet$}
\newcommand{\prt}{$\circ$}
\newcommand{\headrow}[1]{\multicolumn{1}{c}{\adjustbox{angle=45,lap=\width-0.5em}{#1}}}

\definecolor{semi-light-gray}{gray}{0.7}
\definecolor{light-gray}{gray}{0.8}

\usepackage[ruled]{algorithm2e}

\SetAlFnt{\small}
\SetAlCapFnt{\small}
\SetAlCapNameFnt{\small}
\SetAlCapHSkip{0pt}
\IncMargin{-\parindent}

\begin{document}
\title{Server Location Verification and Server Location Pinning:\\ Augmenting TLS Authentication}

\author{\IEEEauthorblockN{AbdelRahman Abdou\IEEEauthorrefmark{1} and P.C. van Oorschot\IEEEauthorrefmark{2}}
\IEEEauthorblockA{School of Computer Science,
Carleton University\\
Ottawa, ON, Canada\\
Email: \IEEEauthorrefmark{1}abdou@scs.carleton.ca,
\IEEEauthorrefmark{2}paulv@scs.carleton.ca,
}}

\newcommand{\manager}{\textsl{Manager}}
\newcommand{\true}{\texttt{true}}
\newcommand{\false}{\texttt{false}}

\maketitle
\begin{abstract}
We introduce the first known mechanism providing realtime server location
verification. Its uses include enhancing server authentication (\eg augmenting
TLS) by enabling browsers to automatically interpret server location
information. We describe the design of this new measurement-based technique,
Server Location Verification (SLV), and evaluate it using PlanetLab. We
explain how SLV is compatible with the increasing trends of geographically
distributed content dissemination over the Internet, without causing any
new interoperability conflicts. Additionally, we introduce the notion of
(verifiable) \emph{server location pinning} within TLS (conceptually similar
to certificate pinning) to support SLV, and evaluate their combined impact
using a server-authentication evaluation framework. The results affirm the
addition of new security benefits to the existing SSL/TLS-based authentication
mechanisms. We implement SLV through a location verification service, the
simplest version of which requires no server-side changes. We also implement
a simple browser extension that interacts seamlessly with the verification
infrastructure to obtain realtime server location-verification results.

\end{abstract}
\IEEEpeerreviewmaketitle

\section{Introduction}
\label{sec:introduction}
\blfootnote{Version: 8$^{th}$ August, 2016.}

Knowledge of a webserver's verified geographic location can provide greater
assurance of the webserver's authenticity, and helps establish the legal
jurisdiction under which the server resides, \eg in case of disputes. The
street address of a domain owner/operator is typically different than
the location of the physical server hosting its content. If a server's
geographic location is verified in realtime, a user-agent (\emph{browser}
henceforth) may, \eg by virtue of a pre-established privacy policy, refrain
from proceeding with a connection knowing that the website is hosted from
a suspicious location, or a jurisdiction lacking solid privacy laws.

Server authentication on the web is primarily achieved using HTTP over
SSL/TLS (or HTTPS) and a distributed PKI, albeit with questionable trust
semantics. A long list of known problems in that architecture have been
identified~\cite{clark2013sok,holz2011ssl}, raising open-ended questions about
the security of the status-quo~\cite{kranch2015upgrading,laurie2014cert}. This
paper reinforces server authentication on the web, by weaving the
server's physical location information into current authentication
mechanisms. This helps mitigate server impersonation attacks such as
phishing~\cite{dhamija2006phishing}, pharming~\cite{li2015pharming}, and the
use of rogue certificates~\cite{vratonjic2013inconvenient} possibly after
a CA compromise.

To achieve such location-based webserver authentication, we successfully
address several challenges. For example, because of the increasing physical
distribution of web content (\eg cloud computing environments, content
distribution networks or CDNs, distributed web-caching and proxy servers, load
balancers, and P2P networks), the traditional server-client model where an
HTTP session is established entirely between a browser and a single physical
server, and content is downloaded only from that server, is becoming less
common. Web content is often fetched from several physical/virtual servers,
possibly not geographically collocated. How can useful location information be
extracted from that context to provide assurance of the domain's authenticity?

Another challenge is the lack of a practical mechanism for realtime server
location verification. IP-based location determination is susceptible
to location spoofing attacks~\cite{MuirPaul}, making it unsuitable for
authentication. Offline location verification, \eg a certification authority
(CA) verifying the server's location at the certificate-issuing time and
binding the issued certificate to the server's location~\cite{kim2013geopki},
does not provide location assurance at time of later interaction with the
server. Additionally, such a solution would require the domain owner to obtain
a certificate for each group of physically collocated webservers, which is
impractical in both cost and complexity for large providers that may have
thousands of servers around the world (\eg Akamai \cite{AkamaiFacts}). On
the other hand, common delay-based IP geolocation schemes work in realtime,
but are susceptible to delay manipulation attacks~\cite{Dude}. Prior to the
work herein, no known realtime server location verification mechanism that
accounts for common adversarial location-forging tactics existed.

To tackle the aforementioned challenges, we introduce Server Location
Verification (SLV)---a measurement-based realtime server location verification
mechanism. Using a network of distributed verifiers\footnote{Verifiers could
be regular servers deployed using VMs, on cloud infrastructure, or physical
servers. Third-parties could provide this in practice, including existing
commercial CDN-providers, or as a non-profit organization.} over the Internet,
the goal of SLV is to verify the geographic location of the first webserver
with which the client has a TCP connection. We explain the design of SLV,
and implement a simple version that requires no server-side changes, which
is thus readily deployable through a browser extension. We also test SLV's
efficacy from a location verification standpoint, and analytically evaluate
its usage as an additional webserver authentication mechanism.

The strong assurance SLV provides to the geographic location of
a server adds a new, beneficial dimension to the current notion
of webserver authentication. Comparing a server's location to its
public key (or certificate), realtime location verification can be
seen as analogous to browser certificate validation. As introduced
herein, \emph{server location pinning} (\eg in the browser) can also
model key pinning~\cite{kranch2015upgrading} to further enhance server
authentication. Browsers can cross-check a server's verified location in a
fashion similar to multipath probing~\cite{wendlandt2008perspectives}. A
list of physical locations where a server is hosting its content from
can be made publicly available for realtime consultation (\cf list of
active certificates~\cite{clark2013sok}). Existing certificate revocation
primitives can be extended to \emph{revoke a location}, \eg if a data centre
was relocated or if content is no longer distributed from a previous mirror.

A domain may legitimately have multiple public keys; primitives such as key
pinning and certificate revocation address that by attempting to recognize
whether an already validated certificate is authentic. Likewise, adopting
analogous primitives based on server location is compatible with legitimately
distributing a domain's content from multiple geographic locations, \eg by
pinning all such locations, or actively revoking obsolete locations.

We make the following contributions to enhance server authentication on
the web:
\begin{itemize}
\itemsep0em
\item conceptualizing for the first time how the idea of physical (geographic)
location of a webserver can be incorporated as an additional dimension to
strengthen server authentication, in a manner compatible with but independent
of current SSL/TLS standards;
\item designing and implementing SLV, a new measurement-based algorithm
for server location verification, which in its simplest form requires no
server-side changes nor human-user interactions, and evaluating its efficacy
through pilot experiments using PlanetLab~\cite{planetlab};
\item augmenting this new mechanism with \emph{browser-based server-location
pinning}---a primitive to enable browsers to establish location-based trust
semantics over time. \end{itemize}

In addition to location-based server authentication, verified server location
may provide evidence for services like cloud providers that their servers are
in a particular country~\cite{peterson2011position}, \eg with more favourable
data privacy laws than others, thus gaining a competitive advantage. Likewise,
e-commerce service providers may benefit from assuring their clients that
payments are processed in a country they expect or are comfortable with.

The rest of this paper is organized as follows. Section~\ref{sec:background}
reviews traffic hijacking tactics, characteristics of Internet
delay measurements, and the distributed nature of fetching web
content. Section~\ref{sec:threatmodel} defines the threat model. SLV is
explained in Section~\ref{sec:methodology}, followed by server-location
pinning in Section~\ref{sec:pinning}. Section~\ref{sec:results} empirically
tests a prototype implementation of SLV, and evaluates the presented
location verification primitives using a server-authentication evaluation
framework. Further discussion is given in Section~\ref{sec:discussion}. We
review related work in Section~\ref{sec:relatedwork} and conclude in
Section~\ref{sec:conclusion}.

\definecolor{light-gray}{gray}{0.95}
\lstset{   backgroundcolor=\color{light-gray},
basicstyle=\scriptsize\ttfamily,  breakatwhitespace=false,
breaklines=true,			captionpos=b,
commentstyle=\color{ForestGreen},	extendedchars=true,
frame=single,				keepspaces=true,
keywordstyle=\color{blue},		language=HTML,
numbersep=5pt,				  numberstyle=\tiny\color{gray},
stepnumber=1,			     rulecolor=\color{black},
showspaces=false,			showstringspaces=false,
showtabs=false,				tabsize=2,
escapeinside={(*@}{@*)},
  stringstyle=\color{orange},
  showstringspaces=false,
}

\section{Background}
\label{sec:background}

This section reviews Internet traffic hijacking mechanisms, the role of timing
measurements in location inference, and mechanisms of content distribution
over the Internet. Readers familiar with this background can proceed to
Section~\ref{sec:threatmodel}.

\subsection{Traffic Hijacking: A Network Perspective}
\label{sec:traffichijack}

Web traffic hijacking is an attack whereby the adversary impersonates the
authentic domain, directing users' requests to a machine under the adversary's
control rather than one under the control of the domain owner.

\textbf{Hijacking at different levels.} Starting by the user initiating a
connection to a domain over the Internet, and moving down the TCP/IP protocol
stack, traffic hijacking could be mounted at every point where a new network
addressing scheme identifying the intended destination is introduced. Such
identifiers include the domain name, IP address, MAC address, and the switch
port that a machine is physically connected to. Note that at higher layers
of the protocol stack, the notion of an \emph{Internet domain} is abstracted,
and can be viewed as a single entity. At lower layers, that entity can become
more distributed across multiple physical or virtual machines. References
to an authentic/intended webserver or machine in what follows denote any
such physical or virtual machines designated by the domain owner to store
and offer the domain's services/content over the network.

Misleading the user to visit a different (visually similar or disguised) domain
name than the intended one is phishing~\cite{dhamija2006phishing}. Because
the identifier here differs at the highest addressing scheme, subsequent
identifiers, namely the IP address, MAC and port connecting the fraudulent
machine to the network, are expected to be different from that of the
authentic webserver. Similarly, a pharming attack~\cite{karlof2007dynamic}
occurs by misleading the browser-consulted name resolver, which could be at
any level in the DNS hierarchical lookup procedure, to resolve the domain
name to an IP address assigned to the adversary's machine. The requested
domain name is thus equal to the intended one, but the IP address and the
remaining identifiers are different from that of the intended machine(s).

ARP spoofing~\cite{kiravuo2013survey} and BGP
spoofing~\cite{goldberg2014taking} are examples of traffic hijacking, where
an adversary misleads switches or routers respectively about the network
location of the authentic webserver. Both the domain name and IP address of
the fraudulent machine are the same as that of the webserver, but the MAC
address is different.

Finally, after the switch knows the MAC address of the intended destination,
it looks up its MAC table for the physical port number where that machine
is plugged. Poisoning the switch's MAC table~\cite{kiravuo2013survey}
causes the adversary to deceive the switch into forwarding the data to the
physical port where the adversary's machine is connected, thus hijacking
traffic intended to the authentic machine. In such a case, the domain name,
IP and MAC addresses of the fraudulent machine match those of the authentic
one, but the switch port number is different.

This background is used later in the threat model, as summarized in
Table~\ref{sec:threatmodel}. Other on-route hijackings are also possible
with other addressing schemes, such as in the Spanning Tree Protocol
(STP)~\cite{ieeestp}, where switches are assigned \texttt{BridgeIDs}, or by
other injection mechanisms~\cite{quantuminsert}.

\textbf{Hijacking versus MitM.} Once traffic is hijacked, the adversary
may itself open another back-end connection to the authentic domain as a
regular user, to present the actual user with seemingly authentic responses
and thus avoid exposure. The adversary thus becomes a Man-in-the-Middle
(MitM)~\cite{ornaghi2003man}, relaying traffic between the user and the
intended domain. Our work herein addresses traffic hijacking in general,
whether it is a \emph{hijack-and-host} or \emph{hijack-and-relay} (MitM).

\textbf{The role of SSL/TLS.} Regardless of where in the network traffic
hijacking occurs, HTTPS using SSL/TLS with a browser-trusted certification
authority (CA) is intended to give assurances about the identity of an
authentic domain, aiming to prevent the adversary from \emph{successfully
impersonating} the authentic domain/webserver. Such successful impersonation
requires not only traffic hijacking, but also defeating TLS protection
mechanisms.

To successfully impersonate an HTTPS-enabled domain, the adversary either
needs to hijack traffic at the highest addressing level---phishing---or at
lower levels, which would also require other actions such as compromising
a browser-trusted CA to bind the domain name to the adversary's private
key, compromising the authentic domain's private key, or downgrading
from HTTPS to HTTP during the connection establishment time, \ie SSL
stripping~\cite{marlinspike2009more}.

While phishing should technically be the easiest to detect since all
addresses identifying the adversary's machine differ from the authentic one,
it remains effective as it relies on social engineering rather than technical
manipulations.

For hijacking traffic at lower levels (as noted above), the Internet's open PKI
system is subject to a single point of failure; a single CA compromise could
jeopardize the security of the entire system~\cite{clark2013sok}. As such, the
system is at most as secure as the weakest CA. Various enhancing primitives
have been proposed, such as certificate pinning~\cite{kranch2015upgrading}
and multipath probing~\cite{wendlandt2008perspectives}, but these aim to
strengthen the current PKI system. In contrast, server location verification
operates orthogonally as an independent webserver authentication dimension.

Other than a CA compromise, previous literature reports domain operators
sharing their private keys among other constituents~\cite{liang2014https},
which corrupts the system's key mechanism of identity assurance. An adversary
with access to the domain's private key need not compromise any CA to mount
a successful impersonation attack; this is undetectable by primitives such
as key pinning.

Next, we review characteristics of Internet delays, and their relationship
to geographic locations over the Internet.

\subsection{Timing-based Measurements}
\label{sec:background:cpv}

Literature over the past decade confirms a strong
correlation between Internet delays and geographic
distances~\cite{GeoTrackGeoPing,streetlevel,Dong201285,landa2013measuring}.
Although network routes are subject to many conditions that may
impede such a correlation, like route circuitousness~\cite{Octant}
and delay spikes due to possible network congestion, the strong
correlation remains~\cite{landa2013large}. This is usually
attributed to constantly improving network connectivity and bandwidth
availability~\cite{Constrainbased}.

Many networking applications have leveraged this
correlation to achieve accurate IP geolocation over the
Internet~\cite{laki2011spotter,LearningBased,geoweight}. A common approach
is to derive functions that map delays to distances based on observing
various network characteristics (topology, latency, \etc). The function
is then used to map delays measured between multiple vantage points
(with known locations) and the target IP address to geographic distances,
thus constraining the region where the machine assigned that IP address
is physically present. Measurement-based location techniques can achieve
high accuracy (\eg a few hundred meters~\cite{streetlevel}), for inferring
geographic information from network measurements. CPV~\cite{cpvtdsc} (see
Section~\ref{sec:relatedwork}) was the first measurement-based technique to
verify \emph{client} location assertions, addressing an adversarial client
aiming to evade~\cite{MuirPaul} geolocation or manipulate those techniques
to its favour~\cite{Dude}.

\subsection{Fetching Web Content}
\label{sec:background:fetching}

We review common methods used for dissemination and delivery of web content.

{\bf Content Distribution Networks (CDN).} A CDN is a network of caching
servers used to distribute web content efficiently. CDNs, which have become
quite popular, aim to offload the effort of managing and distributing content
at large scale from the content owner. Different techniques are used for
managing content replication and redirecting browsers to the appropriate
CDN surrogate server.

Liang \etal~\cite{liang2014https} note two common practices for browser
redirection. The first rewrites the URLs of objects (scripts, images, {\it
etc}) to point to their location on the appropriate CDN server, \eg using
the \texttt{src} HTML attribute. For example, to instruct the browser to
fetch \texttt{image.gif} from the CDN server, the webserver uses:
\begin{lstlisting}
<img src="http://www.cdn-server.com/image.gif">
\end{lstlisting}
instead of
\begin{lstlisting}
<img src="some-local-directory/image.gif">
\end{lstlisting}
The second practice resolves the website's domain name to the IP address of
the respective CDN server, achieved either by a DNS server under the CDN's
administration configured as the authoritative name server for the original
website, or by the website's DNS server itself.
In the first practice, the browser establishes HTTP(S)/TCP connections with
the original server first, and then with the CDN surrogate server; in the
second, the browser only contacts the CDN server without the need to contact
the original server.

{\bf Caching and proxy servers.} A caching server, sitting in the middle of
the connection between the browser and the original webserver, terminates
the TCP connection intended between client and webserver, and re-initiates
another one with the webserver. When the caching server receives an HTTP GET
request to a cached object, it sends a conditional GET request to the original
server that includes the header line \texttt{If-modified-since} specifying the
date/version of the cached object. The server either responds with \texttt{304
Not Modified}, or with the requested object if the cached version is stale.

Caching servers can be set up at any point along the communication path between
client and webserver. For example, the network administrator could set up a
caching server, and route network traffic to it to reduce external network
usage. An ISP could set-up caching servers to manage network congestion,
and the website operator could also set up caching servers to reduce load
on the main server.

A non-caching proxy is sometimes also used, \eg for privacy purposes; the
TCP termination hides the client's IP address from the webserver. If the
client configures its local machine to use a remote proxy, outbound packets
have the proxy's IP address as their destination.

{\bf Other schemes.} Other content distribution schemes and legitimate browser
(re)directs/pointers also exist, such as browser-based ads, collocated
load balancers, fast-flux servers (see Section~\ref{sec:relatedwork}),
authentication servers and P2P networks. These operate largely similar to
the methods reviewed above; we omit further discussion for space reasons.

\section{Threat Model and Assumptions}
\label{sec:threatmodel}

{\bf Adversary's objective. } The threat model assumes an adversary aiming
to impersonate a webserver by hijacking its traffic. The adversary's typical
goals are eavesdropping, or stealing a user's authentication credentials.

{\bf Summary of hijacking mechanisms. } To define the scope of traffic
hijacking mechanisms (see Section~\ref{sec:traffichijack}) included in the
threat model, Table~\ref{table:threats} classifies them by the subset of
adversary's machine identifiers that would be equal to that of the authentic
machine in each mechanism. Identifiers include the \emph{Domain Name},
\emph{IP Address}, \emph{MAC Address}, and \emph{Switch port number} the
machine is connected to (note that Table~\ref{table:threats} is also used
in Section~\ref{sec:results}).

The hijacking level dictates whether an adversary needs to be on-route
between the user and the intended destination. On-route hijacking, \eg ARP
spoofing and MAC table poisoning, need not necessarily be mounted at the
destination network where the intended machine is connected; it may occur
at any intermediate network along the route. If hijacking is mounted at an
intermediate network, the destination IP or MAC addresses of the fraudulent
machine would be equal to that of the router or switch respectively of the next
hop along the route. (Note that the \emph{levels} in Table~\ref{table:threats}
differ from the five TCP/IP \emph{layers} of the protocol stack.)

On-route network hijacking requires the adversary to locally place itself
in one of the intermediate networks, possibly by compromising a host or
switch already part of that network. Rows 1-3 in Table~\ref{table:threats}
do not have that requirement.

\newcommand{\thin}{0.3cm}
\newcommand{\numlines}{1}

\newcommand{\W}{{\tiny $\blacksquare$}}
\newcommand{\w}{{\scriptsize $\bigstar$}}
\newcommand{\D}{{\tiny $\square$}}

\newcommand\circledmark[1][red]{  \ooalign{    \hidewidth
    \kern0.65ex\raisebox{-0.9ex}{\scalebox{3}{\textcolor{#1}{\textbullet}}}
    \hidewidth\cr
    $\checkmark$\cr
  }}
\newcommand{\tick}{}
\newcommand{\tickI}{\hspace{-5pt}$\checkmark$}
\newcommand{\tickIR}{$\checkmark$}

\begin{table}
\centering
\caption{Levels of traffic hijacking, and whether each can affect a local and/or
a global set of clients. A check-mark ( $\checkmark$ ) means the respective
traffic hijacking is included in the threat model.}

\scalebox{0.98}{
\begin{tabular}{@{}crl|c@{}c@{}|@{}R{\thin}@{}R{\thin}@{}R{\thin}@{}R{\thin}@{}}

&&&&&\multicolumn{4}{c}{Identifier}\\\cline{6-9}
\rotl{On-route hijacking?}&Level	&	Example hijacking
&\rott{\scriptsize Local hijacking considered?}&\rots{\scriptsize Global
hijacking considered?}&\rott{\scriptsize Domain Name}&\rott{\scriptsize IP
Address}&\rott{\scriptsize MAC Address}&\rott{\scriptsize Switch port}
\\\hline

\multirow{3}{*}{No}
&\multicolumn{1}{|r}{\emph{(Human)}}	&	Phishing	&\tickI
&\tickIR &\D&\D&\D&\D	\\
&\multicolumn{1}{|r}{Application}		&	Pharming	&\tickI
&\tickIR &\W&\D&\D&\D	\\
&\multicolumn{1}{|r}{AS}				&	BGP
spoofing	&\tick &\tickIR &\W&\W&\D&\D	\\\hline

\multirow{2}{*}{Yes}&\multicolumn{1}{|r}{Network}	&	ARP spoofing
&\tick &\tickIR &\w&\w&\D&\D	\\
&\multicolumn{1}{|r}{Link}		&	MAC table poisoning
&\tick &\tickIR &\w&\w&\w&\D	\\\hline

---&\multicolumn{1}{|r}{Physical}	&	\emph{no network hijacking}
&\hspace{-4pt}-&\hspace{-4pt}-&\W&\W&\W&\W	\\

\hline
\end{tabular}
}
\scriptsize
The \emph{Identifier} assigned to the hijacker's machine is either different
from (\D), or same as (\W) that of the intended destination or the next hop
machine along the route (\w).
If the on-route hijacking occurs at the final destination network, then \w\
are replaced with \W \label{table:threats}
\end{table}

{\bf Local versus global effect. } Note that each hijacking mechanisms
in Table~\ref{table:threats} may affect either a global or a local set
of clients. For example, an adversary compromising the intended domain's
authoritative DNS server itself, can mount pharming attacks on essentially
all clients visiting that domain; on the other hand, spoofing local DNS
resolutions affects only a local subset of clients. If spoofed BGP prefix
announcements propagate to large portions of the Internet, they result in
a global effect. Otherwise, their effect is local to the set of affected
networks. For on-route hijackings, the closer they are to the network of
the intended destination, the more global their effect. For example, ARP
spoofing and MAC table poisoning mounted within the local network of the
intended destination will affect almost all visiting clients.

{\bf Adversarial Capabilities Assumed. } The adversary is capable of hijacking
traffic such that its fraudulent machine's IP address differs from that of the
authentic webserver, \ie Table~\ref{table:threats}'s first two levels. This
covers a set of local hijacking, including local pharming attacks.

For attacks where the fraudulent machine's IP address is equal to
that of the authentic destination (\ie on the AS, Network and Link
levels of Table~\ref{table:threats}), the threat models includes
global hijacking attacks. This includes the 2008 Pakistani Telecom
incident~\cite{pakistaniincident} and that of China Telecom in
2010~\cite{hiran2013characterizing}. The threat model also encompasses
low-level network hijacks, like ARP spoofing and MAC table poisoning, mounted
within the local network of the intended destination as it will affect almost
all visiting clients---see Table~\ref{table:threats}.

The threat model excludes hijacking mechanisms that satisfy the following two
conditions together (the three cells without $\checkmark$ under \emph{Local
effect} in Table~\ref{table:threats}): (1) they affect only a local subset
of clients and (2) they are conducted on a level where the IP address of
the fraudulent machine is equal to that of the intended one. An adversary
mounting this class of hijacking can bypass location verification because
the selected verifiers (those used to verify a location, as explained in
Section~\ref{sec:methodology}) may not be affected by that locally-impactful
traffic hijacking. Thus, they might end up verifying the location of the
authentic machine, as identified by the client-submitted IP address, vs.\
the fraudulent one.

The mechanisms presented herein can provide further assurance to a webserver's
identity in the absence of HTTPS. The threat model thus assumes an adversary
that may or may not compromise the domain's SSL/TLS server private key, or
issue a fraudulent certificate possibly by compromising a browser-trusted
CA. The model also addresses the case of an adversary that can mount SSL
stripping attacks~\cite{marlinspike2009more} to downgrade a connection, and
generally other SSL/TLS-related attacks. It is assumed that the verifiers
are trusted to carry out and report delay measurements honestly.

\section{Server Location Verification}
\label{sec:methodology}

This section introduces the measurement-based Server Location Verification
(SLV) technique. SLV leverages generic delay measurement guidelines from
previous literature to infer location information from timing analysis
\cite{GeoTrackGeoPing,Constrainbased,laki2011spotter}. It is custom-designed
to verify Internet webserver locations.

{\bf Location Assertion.}
There exists no standard mechanism to enable a webserver to assert its
geographic location to a browser, \eg no standard HTTP headers convey
that information. For deployability benefits, we rely on IP geolocation
databases~\cite{huffaker2011geocompare} for location assertions (these
assertions will be verified by SLV), thus requiring no server-side changes
nor any additional server involvement. These databases may occasionally
have outdated or coarse-grained IP address location information; webserver
provision of an accurate location assertion, \eg through HTTP headers, would
thus be beneficial but comes at the cost of server-side changes. The design
of SLV allows alternate more reliable sources of location assertions.

{\bf Location Verification.}
SLV is designed to verify the geographic location of the first machine the
browser establishes a TCP connection with. This is the machine assigned
the IP address resulting from the domain name resolution. This enables SLV
to address all pharming attacks, regardless of where in the hierarchical
lookup procedure an adversary may spoof the name resolution; SLV itself does
not contact any DNS systems for name resolutions. If the browser receives a
spoofed IP address via DNS due to a pharming attack, that IP address is the
one passed to SLV for verification. Accordingly, a fraudulent IP address from
a local pharming attack would be presented to SLV for location verification.

\subsection{Architecture and Algorithm}

\begin{figure}
\centering
\includegraphics[scale=0.4]{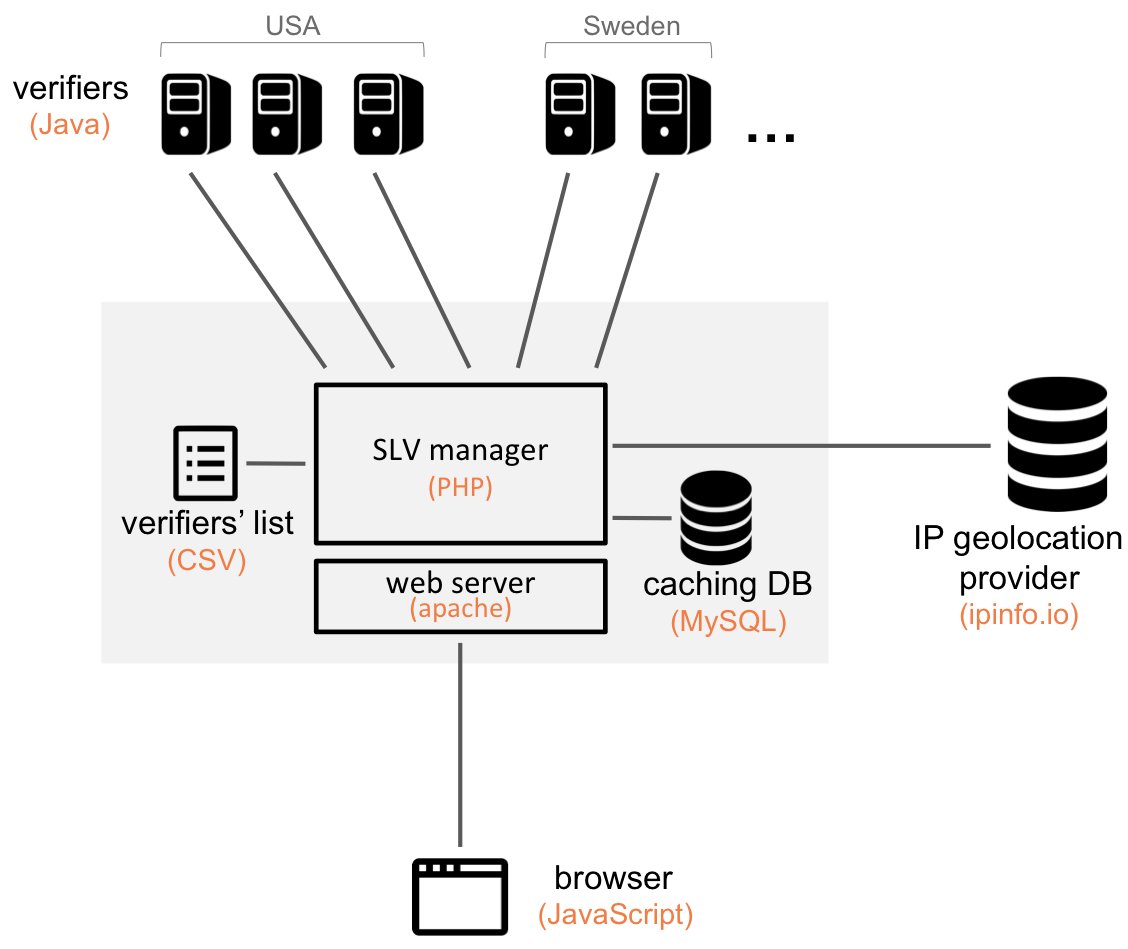}
\caption{System architecture. Parenthesized terms correspond to implemented
prototype (Section~\ref{subsec:implementation})}\label{fig:architecture}
\end{figure}

The system's architecture is shown in Fig.~\ref{fig:architecture}. The SLV
\manager\ is an independent server acting as an interface between a browser
and the \emph{verifiers}. A verifier is a machine, \eg a virtual private
or a cloud-based server, used to measure Internet delays to a webserver as
instructed by the \manager. The \manager\ itself runs on top of a webserver,
and has access to a list of distributed verifiers and their geographic
locations. For efficiency, the \manager\ caches location verification results;
entries for a given IP address are cached for a configurable period (\eg a
few hours), before expiring.

Algorithm \ref{algo:main} details the location verification process (see
Table~\ref{table:datatype:main} for data structures used). When a user
visits a website, the browser sends the resolved IP address (the input
$adrs$ in Algorithm~\ref{algo:main}) of the website to the \manager. By the
\emph{locate()} function in line~\ref{line:assert}, the \manager\ obtains
the best available assertion of the server location (the simplest case may
involve using the server's IP address, \eg using IP-to-location mapping
databases~\cite{Tabulation}). The returned result serves as an unverified
assertion of the server's location. The \manager\ then checks for a cached
verification result of that address (see the \texttt{result} structure
in Table~\ref{table:datatype:main}), and returns it to the browser if the
location corresponding to the IP address has not changed since caching time
(\eg the IP address was not assigned to another machine somewhere else).

If no entry was cached, or the cached location is not equal to the newly
asserted one, the \manager\ begins location verification by selecting
three verifiers geographically encompassing the asserted location
(line~\ref{line:selectv}).
Each verifier measures the round-trip time (RTT) to the target
webserver and to the other two verifiers (line~\ref{line:RTT}). RTTs
are not measured using standard ICMP-based tools; this avoids QoS
and routing policies at intermediate ASes from artificially delaying
or dropping probing messages, and other known problems of such
techniques~\cite{Dude,Augustin:2006:ATA:1177080.1177100}. Instead, the
verifiers measure RTTs over the application layer by initiating a TCP
connection to the target server like a regular web client, and calculating
the RTT from the \texttt{SYN}--\texttt{SYNACK} handshake. The verifiers
conduct several RTT measurements, and send the smallest back to the \manager.

Due to last-mile delays~\cite{Octant}, the delay-to-distance ratio is
inflated near the edge networks of two communicating parties over the
Internet. Such inflation occurs four times while measuring the RTTs between
a pair of verifiers and the webserver (\ie twice between each verifier and
the webserver). On the other hand, RTTs are inflated only twice when measured
directly between a pair of verifiers. As such, we filter out the inflation
factor by subtracting a value, $\lambda$, from the three RTT measurements
between each verifier and the webserver. In practice, the value of $\lambda$
may be calibrated in realtime, \eg as the average RTT between all verifiers
in the region and their network gateway. Extensive delay analysis in previous
literature found the network edge causes a delay inflation equivalent to
$\sim$5ms~\cite{streetlevel}. We use this value in our prototype implementation
(see Section~\ref{subsec:implementation}).

After subtracting $\lambda$, the \manager\ stores the delay information in
a two-dimensional array $D$, in line~\ref{line:RTT}, such that $D[v][i]$
is verifier $v$'s measured RTT between itself and entity $i$; $i$ is either
another verifier or the target webserver.

\newcommand{\mynl}{\hspace{8pt}\nl}

\begin{algorithm}
\DontPrintSemicolon
{\bf Inputs}\\
\begin{tabular}{ll}
$\mathbb{C}[.]$:	& An array of \texttt{result} cached at the \manager\\
$adrs$:				& A \texttt{string} of the server's IP address
\end{tabular}\\
\KwOut{\\
\begin{tabular}{ll}
$res$:				&A structure of verification \texttt{result}\\
\end{tabular}
}
\Begin
{
	\mynl Declare $ip$ of type \texttt{IP\_info}\;
	\mynl Declare $res$ of type \texttt{result}\;

	\mynl $ip.value:=adrs$\;
	\mynl \label{line:assert}$ip.loc:=$ locate($adrs$)\;

	\mynl \If{$ip.value$ \emph{\bf exists in} $\mathbb{C}$}{
		\mynl $res:=\mathbb{C}[ip.value]$\;
		\mynl \If{$res.ip.loc=ip.loc$}{
			\mynl \Return $res$\;
		}
	}
	\mynl $res.ip:=ip$\;
	\mynl $res.when\_veri:=$ local time at the \manager\;

	\mynl Declare $circ$ of type \texttt{circle}\;
	\mynl $T:=$ the set of triangles geographically encompassing $ip.loc$\;
	\mynl \label{line:bigloop}\ForEach{$t$ \emph{\bf in} $T$}{
		\mynl \label{line:selectv}$V:=$ the three verifiers determining
		$t$\;
		\mynl \label{line:RTT}$D:=$ RTTs between the
		verifiers in $V$ and $ip$\;
				\mynl \ForEach{\emph{verifier pair} $[v1,v2]$
				\emph{\bf in} $V$}{
			\mynl \label{line:centre}$circ.centre:=$ mid\_point($v1$,
			$v2$)\;
			\mynl \label{line:radius}$circ.radius:=$ distance($v1$,
			$v2$)/2\;
			\mynl \If{\emph{RTTs {\bf in}} $D$ \emph{\bf indicate}
			$ip$ \emph{\bf inside} $circ$}{
				\mynl $res.veri\_passed:=$ \texttt{true}\;
				\mynl $res.region:=circ$\;
				\mynl $\mathbb{C}:=\mathbb{C}\cup res$\;
				\label{line:circveri}\mynl \Return $res$\;
			}
		}
	}

	\mynl $res.veri\_passed:=$ \texttt{false}\;
	\mynl \label{line:null}$res.region:=$ \texttt{null}\;
	\mynl $\mathbb{C}:=\mathbb{C}\cup res$\;
	\mynl \Return $res$\;
}
\caption{Location verification run by the \manager. See
Table~\ref{table:datatype:main} for data types, and inline for explanation.}
\label{algo:main}
\end{algorithm}

\begin{table}
\centering
\caption{Structure of exchanged messages}
\begin{tabular}{@{}c|ccL{2cm}}\hline

\hline
Struct&Attribute	&	Data Type	&	Description\\

\hline
\multirow{4}{*}{\texttt{result}}
&$ip$			&	\texttt{IP\_info}	&	 See
\texttt{IP\_info} below\\
&$veri\_passed$ &	\texttt{boolean}	&	 Verification
result\\
&$region$		&	\texttt{circle}		&	 Verification
granularity\\
&$when\_veri$		&	\texttt{timestamp}	&	 Date and
time of last verification\\

\hline
\multirow{2}{*}{\texttt{IP\_info}}
&$value$		&	\texttt{string} &	 IP Address \eg
``1.2.3.4"\\
&$loc$		&	\texttt{location}	&	 See \texttt{location}
below \\

\hline
\multirow{2}{*}{\texttt{circle}}
&$centre$		&	\texttt{location}	&	 Centre of
a circle (See \texttt{location} below)\\
&$radius$		&	\texttt{double} &	 Radius of the circle
(\eg in km)\\

\hline
\multirow{2}{*}{\texttt{location}}
&$lat$		&	\texttt{double}		&	 Latitude\\
&$lon$		&	\texttt{double}		&	 Longitude\\\hline

\hline
\end{tabular}
\label{table:datatype:main}
\end{table}

\begin{figure}
\centering
\includegraphics[scale=0.3]{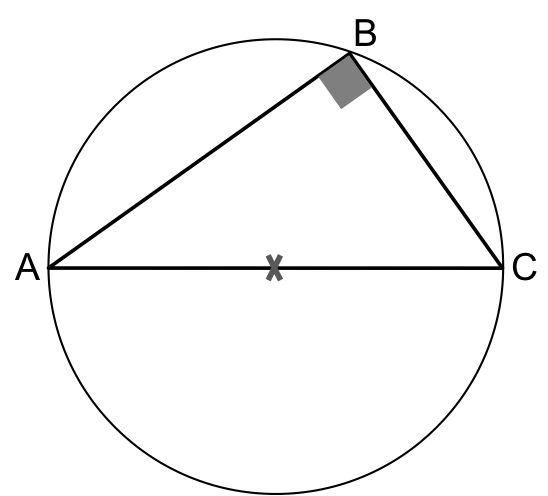}
\caption{Thales' theorem~\protect\cite{book:geometry} is used herein for
server location verification. It states that for an inscribed triangle
(as shown), the angle $\angle ABC=90^{\circ}$}
\label{fig:thale:theory}
\end{figure}

{\bf Geometric Verification. }
By Thales' theorem~\cite{book:geometry}, an inscribed triangle with one side
being the diameter of the prescribing circle is a right-angled triangle, with
the diameter its diagonal (Fig.~\ref{fig:thale:theory}). SLV verifies location
assertions using this, where times are treated as distances consistently for
a given pair of verifiers at a given instant in time; the asserted location is
positively verified if, for any of the three pairs of the selected verifiers,
the sum of the squared RTTs between each verifier and the target webserver
does not exceed the square of the average RTT between the pair, \ie
\begin{equation}
\label{equ:main}
(D[v1][ip])^2 + (D[v2][ip])^2 \leq \left(\frac{D[v1][v2]+D[v2][v1]}{2}\right)^2
\end{equation}

\begin{figure}
\centering
\includegraphics[scale=0.3]{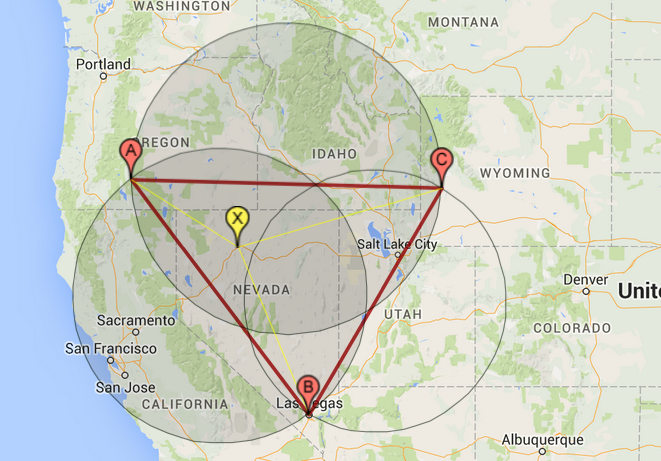}\caption{Example using
Thale's theorem  (map data: Google, INEGI). Each pair of verifiers determines
a unique circle whose centre is the midpoint between both verifiers, and radius
is half the distance between them. Because server $X$ is geographically inside
the circle determined by the pair $[A,B]$, it follows from Thales' theorem
(see Fig.~\ref{fig:thale:theory}) that $AX^2+XB^2\leq AB^2$. Similarly for
the circle determined by $[A,C]$.}
\label{fig:thale:us}
\end{figure}

\noindent Figure~\ref{fig:thale:us} shows an example webserver encapsulated
by a triangle determined by three verifiers. The webserver is inside the two
circles whose diameters are delimited by verifiers $[A,B]$ and $[A,C]$. If
the measured RTTs of one of those pairs of verifiers support the webserver's
presence inside the respective circle, \ie  inequality (\ref{equ:main}) above
holds, this circle becomes the verification granularity. The \manager\ then
signs and sends the verification response to the browser, along with the centre
and radius of the circle (lines~\ref{line:centre} and \ref{line:radius}),
and caches the result for that IP address. If the verification result is
negative for all three circles, a new encompassing triangle determined
by three different verifiers is selected. That process is repeated
(line~\ref{line:bigloop}) until (a) the location is positively verified,
or (b) the verifiers of all triangles (or a sufficient subset thereof)
encompassing the asserted location are exhausted. In our experiments (see
Section~\ref{sec:results}), typically about four triangles suffice for this
test. In (b), a negative verification result is returned, and the granularity
field ($region$) is set to \texttt{null} (line~\ref{line:null}). The
justification for a positive verification from a single triangle being
deemed sufficient to pass location verification is that RTT delays have a
lower bound restricted by the spanned geographic distance (data flows in
fibre at two-thirds the speed of light~\cite{speedoflight}).

{\bf Caching. }
For efficiency, the system employs two layers of caching: one at the browser
and another at the \manager. The former is per browser instance, and is cleared
when the browser process terminates. Browser caching is useful as it helps
when the user switches between tabs or refreshes a page. A cached entry is as
simple as a tuple of $<$IP address, verification decision$>$. Note that results
are cached by IP address, not domain name. Thus, a page refresh that results
in a different non-cached IP address upon resolving its domain causes the
browser to resend the new IP address for verification. This makes the browser
check the webserver's location if the domain's resolved IP address changes.

The caching at the \manager\ is more persistent. A cached entry is also
addressable by IP address, but formatted differently as: $<$IP address,
asserted location, verification date and time, verification result, centre,
radius$>$. The last two entries are the centre and radius of the circle
delineating the verification granularity, if the location was positively
verified, and \texttt{null} otherwise. Because the \manager's caching is
centralized, \ie relative to the browsers' local caching, a browser is likely
to get an instant response from the \manager\ as more verification requests
to the \manager\ are made by an increasing number of relying clients.

\subsection{Implementation}
\label{subsec:implementation}
In our prototype implementation (see Fig.~\ref{fig:architecture}), we used
Apache as our webserver, PHP for the \manager, MySQL for the \manager's
caching, and Java for the verifiers. The communication between \manager\
and verifiers uses standard TCP sockets. The \manager\ learns about the $m$
available verifiers using a simple csv-formatted file with $m$ lines and 3
entries per line: the verifier's IP address, its geographic latitude, and
longitude. We used $m=20$ PlanetLab nodes as verifiers in our testing (see
Section~\ref{sec:results}), which were situated in the US. For IP address
location lookups, we used the ipinfo\footnote{\url{http://ipinfo.io/}} DB.

{\bf Browser extension.} On the browser side, we implemented a Mozilla Firefox
extension to submit the website's IP address to the \manager, and process a
response. We used jQuery to receive verification results from the \manager\
asynchronously. Verification responses are cached locally by the browser,
independent of the caching layer on the \manager, to avoid re-consulting the
\manager\ on page refreshes and tab-switches. The extension also implements
server location pinning, as explained below.

\section{Pinning of Server Location}
\label{sec:pinning}

We introduce the idea of (verifiable) \emph{server location pinning},
following the idea of certificate pinning~\cite{kranch2015upgrading}; as
such, we first quickly review certificate pinning. Certificate pinning is
one approach\footnote{Other items could be pinned, such as the public key
value.} introduced in the SSL/TLS ecosystem to reduce the user's required
interpretation of cues and decision making, shifting that responsibility to the
browser. A website's certificate is \emph{pinned} (or saved) in the browser,
such that future certificates presented by the website are cross-checked
against the saved one, with non-matches typically raising suspicion. A domain's
certificate can be pinned in the browser or in DNS records~\cite{rfc7671}. A
browser can pin certificates (1) automatically as the user browses the web,
(2) when instructed by the server, \eg through HPKP~\cite{rfc7469}, or (3)
when preloaded with pinned certificates. Note that non-DNS based methods
(1) and (3) do not require server-side changes, and can thus be immediately
deployed through browser extensions.

We put forward the principle of server location pinning. A set of
expected server locations (\eg geo-coordinates) is saved locally (by one
of several means explained below) by the user's browser for future cross
checking. When a website is then visited and its location is verified (see
Section~\ref{sec:methodology}), the browser checks if the verified location
falls within any of the pre-pinned regions. The action upon a failed check
can then be handled by, \eg a pre-specified policy. Such a policy might
handle three possible pinning-validation outcomes, as we explain below:
\emph{Critical}, \emph{Suspicious} and \emph{Unsuspicious}. The policy
mechanism is outside of the scope herein. However, a simple intuitive policy
could instruct the browser to refrain from any connection with login forms
or financial transactions in case of \emph{Critical} or \emph{Suspicious}
outcomes; in the absence of login forms and financial transactions, the
browser drops the connection only in the case of a \emph{Critical} outcome.

Since the geographic locations where a website is hosted from could change
frequently for some websites (\eg due to different content distribution
architectures as explained in Section \ref{sec:background:fetching}),
server-side cooperation can provide the benefit of dictating which
geographic locations should be pinned. This could be, for example, in
the form of (1) a publicly queryable set of websites and their locations,
which can also provide the benefit of quick location updates; (2) realtime
location pinning instructions possibly in the form of HTTP headers created
by the webserver itself; and (3) incorporating server location updates into
DNS.\footnote{DNS location records (\texttt{LOC}) were initially proposed
as a means of disseminating IP-address location information \cite{rfc1876}.}
These examples respectively can be viewed as conceptually analogous to trusted
directories or Online Certificate Status Protocol (OCSP) \cite{clark2013sok},
HTTP Public Key Pinning (HPKP) \cite{rfc7469}, and DNS-Based Authentication
of Named Entities (DANE) \cite{rfc7671} in the SSL/TLS ecosystem.

\newcommand{\hwarning}{$outcome:=$\emph{Critical}}
\newcommand{\swarning}{$outcome:=$\emph{Suspicious}}

\begin{algorithm}
\DontPrintSemicolon
{\bf Inputs}\\
\begin{tabular}{lL{7cm}}
$\mathbb{P}[.]$:  &	An array of pinned domain locations\\
$d$: &	Domain in question\\
$r$: &	Verification \texttt{result} of Algorithm~\ref{algo:main}\\
\end{tabular}\\
{\bf Goal}\\
\ \ To enable the browser to establish location-based trust semantics
over time.\\
\Begin
{
	\mynl $outcome:=$\emph{Unsuspicious}\;
	\mynl \If{$d$ \emph{\bf exists in} $\mathbb{P}$}{
		\mynl $pin:=\mathbb{P}[d]$\;
		\mynl \If{$r.veri\_passed=$ \emph{\texttt{false}}}{
			\mynl \hwarning\;
		}
		\mynl \Else{
			\mynl \label{twoconditionsmet} $found:=\false$\;
			\mynl \ForEach{$region$ {\bf in} $pin.regs$}{
				\mynl $l:=$ dist($r.ip.loc$, $region.centre$)\;
				\mynl \If{$l \leq region.radius$}{
					\mynl $found:=$ \true\;
					\mynl \If{$r.ip.value$ \emph{\bf exists
					in} $pin.ips$}{
						\mynl
						$pin.ips[r.ip.value].loc:=r.ip.loc$\;
						}
					\mynl \Else{
						\mynl $pin.ips:=pin.ips\cup
						r.ip$\;
					}
					\mynl $pin.when\_veri:=r.when\_veri$\;
					\mynl \texttt{break}\;
				}
			}
			\mynl \If{$found=$ \emph{\false}}{
				\mynl \If{\emph{\bf size of }
				$pins.regs<pin.rmax$}{
					\mynl \label{addregion}
					$pin.ver\_regs:=$\\ \ \ \
					\ \ \ \ \ \ $pin.ver\_regs\cup r.ver\_regs$\;
				}
				\mynl \Else{
					\mynl \hwarning\;
				}
			}
		}
	}
	\mynl \label{nonfound} \ElseIf{$r.veri\_passed=$ \emph{\texttt{true}}}{
		\mynl Declare $x$ as a pinning struct (see
		Table~\ref{table:datatype:pinning})\;
		\mynl $x.name:=d$\;
		\mynl $x.ips:=pin.ips\cup r.ip$\;
		\mynl $x.ver\_regs:=pin.ver\_regs\cup r.ver\_regs$\;
		\mynl $x.when\_veri:=r.ver$\;
		\mynl $x.when\_pin:=$local time at the browser\;
		\mynl $\mathbb{P}:=\mathbb{P}\cup x$\;
	}
	\mynl \Else{
		\mynl \swarning\;
	}
	\mynl \Return{$outcome$}\;
}
\caption{Server location pinning in the browser.}
\label{algo:pinning}
\end{algorithm}

{\bf Location Pinning Algorithm. }
Locations are pinned as an array $\mathbb{P}$ of the data structure shown in
Table~\ref{table:datatype:pinning}. The array is referenced by the domain
name ($name$). The attribute $ips$ is an array of IP addresses that $name$
(\ie domain name) has previously resolved to. $regs$ is an array of geographic
regions, each described as a centre and radius of a circle, where the domain
name was verified to be hosted from. $rmax$ is the upper limit on the number
of allowed server locations (\eg dictated by the domain operator).

\begin{table}
\centering
\caption{Structure of pinned locations for a domain (see
Table~\ref{table:datatype:main})}
\begin{tabular}{@{}lcL{4cm}}\hline

\hline
Attribute	&	Data Type	&	Description\\
\hline
$name$			&	\texttt{string}
&  The domain name\\
$ips[.]$		&	\texttt{IP\_info}
&  An array of the domain's saved IP addresses \\
$ver\_regs[.]$	&	\texttt{circle}					&
An array of the domain's verified regions\\
$rmax$			&	\texttt{Integer}
&  Upper limit on the number of allowed server locations\\
$when\_veri$	&	\texttt{timestamp}			&  Date and
time of last verification\\
$when\_pin$		&	\texttt{timestamp}			&
Date and time of pinning\\
\hline
\end{tabular}

\label{table:datatype:pinning}
\end{table}

Algorithm \ref{algo:pinning} details the server location pinning
mechanism. When a location verification response is received from the SLV
\manager\ (see Section \ref{sec:methodology}), the browser first searches
$\mathbb{P}$ for a previously pinned location entry for the corresponding
domain name. If none is found (line~\ref{nonfound}), the browser either pins
the domain's location if it was verified, or reacts to a \emph{Suspicious}
outcome as specified by the policy if location verification fails. If a
pinned location is found but location verification has failed, it is a
\emph{Critical} outcome.

Assuming the browser had previously pinned server locations for that domain,
and that the domain's IP address is verified (line~\ref{twoconditionsmet}),
the browser checks if the domain's asserted location falls within any of the
pinned regions for that domain. If it does, the browser either updates the IP
address's corresponding stored geographic locations, or if the IP address was
seen for the first time for that domain, adds it to the array of IP addresses
corresponding to the domain name. If the asserted location does not fall within
any of the pinned locations (but was positively verified), the browser adds it
to the pinned domain as a new region only if more regions are allowed for that
domain (line~\ref{addregion}). Otherwise, the new asserted location, despite
being successfully verified, is classified as a \emph{Critical} outcome.

Note this algorithm does not place any restrictions on the number of IP
addresses allowed per domain. The restriction is only on the number of
different geographic regions ($rmax$) where content is initially provided. In
practice, the value $rmax$ might be set and announced by each domain operator.

\section{Evaluation}
\label{sec:results}

We evaluate SLV in two stages. First, we establish the conceptual validity of
the measurement-based location verification technique itself from a networking
perspective, by attempting to verify websites with known locations using
a prototype implementation. Second, we evaluate the benefits of combining
this with server location pinning to augment server authentication mechanisms.

\subsection{Evaluating Measurement-based SLV}

Our pilot testing uses PlanetLab~\cite{planetlab}, employing as verifiers 20
testbed nodes distributed in North America. We measure the false reject (FR)
and false accept (FA) rates when using the described SLV approach to verify
server location assertions. As such, we test SLV by verifying locations
of servers in which we have available ground truth about their geographic
locations. We followed the assumption that university websites are hosted
on-campus~\cite{streetlevel}, thus we can use their posted street addresses
as our approximation for their webserver locations. See below on verifying
this assumption.

Note that regardless of the content-distribution scheme employed in
practice by a website (\cf Section~\ref{sec:background:fetching}),
a browser always downloads content from one or more physical or virtual
server(s). We focus here on SLV's feasibility to provide measurement-based
location assurance to the server currently being contacted (as explained in
Section~\ref{sec:methodology}), whether that server is standalone or part
of a larger distribution network (\eg a CDN). University servers with known
ground-truth locations thus suffice for our evaluation purpose.

\subsubsection{False Rejects}
We randomly selected 94 university/college websites for testing, and excluded
11 of these that simple filtering found to be hosted by a cloud or a CDN,
lacking ground truth knowledge of their true geographic locations. The
filtering involved looking up from public registries, the AS from which
the domain is reachable. As Table~\ref{sec:methodology} shows, two of the
83 remaining domains were falsely rejected; these both fell in a region
deficient in verifiers. In exploring this, we found one involved verifier
(PlanetLab node) which contributed to both FRs was extremely slow, including
in responsiveness to running commands/processes, thus presumably suffering
technical problems. From this initial study, we know how to reduce the FR rate
below 2.4\% (\ie by testing the reliability of verifier nodes in advance),
but reporting this initial result highlights the importance of a responsive
and sufficient verification infrastructure.

\subsubsection{False Accepts}
To evaluate FAs, the SLV \manager\ was manually configured to select
triangles not encapsulating the asserted locations, and far enough away to
have the asserted location outside the three circles determined by each pair
of verifiers, as explained in Section~\ref{sec:methodology}. The expectation
is that the false location assertion will be correctly \emph{rejected}. Four
triangles were randomly chosen for each tested domain. The verifiers
determining each of four triangles must reject presence inside the respective
circles for a reject decision to result; the parameter four was empirically
determined, and is subject to adjustment. Again, domains were chosen randomly
from among those for which we had ground truth knowledge of webserver location.

One hundred domains were chosen in the following manner: 40 in Europe,
20 in eastern Asia, 20 in Latin America, and 20 in Oceania. As summarized
in Table~\ref{sec:methodology}, none of the tested domains was falsely
accepted. The false accept rate of 0\% is not intended to claim perfection,
but rather is an artifact of limited preliminary testing.

\begin{table}
\centering
\caption{Summary of pilot evaluation results.}
\begin{tabular}{@{}lC{1cm}C{1cm}C{1cm}C{1.5cm}@{}}
					&	\rotl{Total}	&
					\rotl{Accepted} &	\rotl{Rejected}
					&	\rotl{FR/FA}\\\hline
True assertions &	83		&	81	&	2	&
2.4\% FR$^{*}$\\
False assertions	&	100	&	0	&	100	&
0\% FA\\\bottomrule
\end{tabular}
\scriptsize
$^{*}$See inline for an explanation of the results
\label{table:results}
\end{table}

\subsection{Evaluation with Server Location Pinning}

Table~\ref{tab:prims} uses a webserver-authentication evaluation
framework almost identical (in columns) to that developed by Clark
\etal~\cite{clark2013sok}. The column headers show the evaluation criteria,
and the rows are the enhancement primitives.

\newcommand{\dimd}{\cellcolor{light-gray}}

\begin{table*}[th!]

\renewcommand{\arraystretch}{1.3}

\caption{Evaluation of location-based primitives to augment webserver
authentication. \full\ denotes the primitive provides the corresponding
benefit (column); \prt\ denotes partial benefit; an empty cell denotes absence
of benefit. The shaded row (indexed 0), which is itself rated compared to
regular HTTP, serves as the baseline for comparatively assessing improvements
or retrogression of the new primitives detailed herein (rows 1-6).}
\label{tab:prims}
\centering
\scalebox{0.9}{
\begin{tabular}{rl|cccccccc|cccc|cccc|ccc|}

\headrow{ } &
\headrow{ } &
\headrow{Detects MitM} &
\headrow{Detects Local MitM} &
\headrow{Protects Client Credential} &
\headrow{Updatable Pins} &				\headrow{Detects
TLS Stripping} &
\headrow{Affirms POST-to-HTTPS} &
\headrow{Responsive Revocation} &		\headrow{Intermediate CAs
Visible} &
\headrow{No New Trusted Entity} &
\headrow{No New Traceability} &
\headrow{Reduces Traceability} &		\headrow{No New Auth'n
Tokens} &
\headrow{No Server-Side Changes} &
\headrow{Deployable without DNSSEC} &
\headrow{No Extra Third Party} &
\headrow{Internet Scalable} &

\headrow{No False-Rejects} &			\headrow{Status Signalled
Completely} &
\headrow{No New User Decisions}  \\ \cline{3-21}
\multicolumn{2}{l|}{}&
\multicolumn{8}{c|}{\textit{Security Properties Offered}}&
\multicolumn{11}{c|}{\textit{Beneficial Properties}} \\ \cline{3-21}

\multicolumn{1}{l}{ }&
\multicolumn{1}{l|}{ \textit{Primitive}}&
\multicolumn{4}{c|}{ A}&
\multicolumn{2}{c|}{ B}&
\multicolumn{2}{c|}{ C}&
\multicolumn{4}{c|}{\textit{Security/Priv}}&
\multicolumn{4}{c|}{\textit{Deployability}}&
\multicolumn{3}{c|}{\textit{Usability}} \\ \hline

\dimd0&\dimd\textsl{Standard HTTPS}	&\dimd\prt	&\dimd\prt
&\dimd\prt	&\dimd	&\dimd	&\dimd	&\dimd	&\dimd	&\dimd &\dimd\full
&\dimd &\dimd &\dimd &\dimd\full &\dimd\full &\dimd\full &\dimd &\dimd
&\dimd \\
\hline

1&SLV without location pinning	&\full	&\prt	&\prt	&		&\prt
&	&		&	&	&	&	&		&\full
&\full	&	&\prt		&	&\full	&	\\
\hline

2&Server loc. pinning (Client History)		&\prt	&\prt	&\prt	&
&	&	&	&		&\full	&\full	&	&\full
&\full	&\full	&\full	&\full		&	&	&	\\
3&Server loc. pinning (Server)				&\prt	&\prt	&\prt
&	&	&	&	&		&\full	&\full	&	&
&	&\full	&\full	&\full		&\full	&	&\full	\\
4&Server loc. pinning (Preloaded)			&\full	&\full	&\full
&\full	&	&	&		&	&\prt	&\full	&	&\full
&\prt	&\full	&\full	&		&\full	&\prt	&\full	\\
5&Server loc. pinning (DNS)				&\full	&\full	&\full
&\full	&	&	&	&		&\prt	&\full	&\full	&
&\prt	&	&\full	&\full		&\full	&\prt	&\full	\\
\hline

6&List of expected locations			&		&
&		&	&	&	&\full	&		&	&
&	&\full		&\full	&\full	&	&\full		&\full
&\full	&\full	\\\hline

\end{tabular}
}
\end{table*}

{\bf Baseline HTTPS.}
To identify new benefits relative to the standard HTTPS defense mechanism,
we first evaluate HTTPS itself as a baseline for comparison. Row 0 was not
required in the work of Clark \etal~\cite{clark2013sok}, which specifically
evaluated SSL/TLS-enhancements. In our row 0, all comparative evaluation
criteria, such as \textit{No New Trusted Entity} and \textit{No Extra Third
Party}, are relative to regular HTTP (non-HTTPS); the other rows are rated
relative to row 0.
HTTPS provides the first three security properties in
Table~\ref{tab:prims} but only partially, in light of the recent
community awareness of critical HTTPS weaknesses and real-world
attacks~\cite{heninger2012mining,fahl2012eve,vratonjic2013inconvenient}. The
attack surface includes CA compromise, SSL stripping, implementation
vulnerabilities, misconfiguration, and reliance on users to make security
decisions.

Basic HTTPS relies on trusting CAs and signed certificates for server
authentication, and thus lacks bullets at \textit{No New Trusted Entity}
and \textit{No New Auth'n Tokens}. While not introducing new traceability
avenues, it does not \textit{reduce} traceability because revocation methods,
such as OCSP responders, are still required for revocation. It requires servers
to obtain certificates, thus lacks \textit{No Server-Side Changes}. Finally,
HTTPS lacks all three usability properties relative to HTTP.

{\bf SLV.}
For the security properties, SLV (row 1) provides both the benefit of detecting
global MitM attacks, regardless of how the adversary hijacks traffic (recall
Section~\ref{sec:traffichijack}), and of detecting a subset of local hijacks
(column 2 in Table~\ref{tab:prims}), including local pharming attacks. As
noted in Section~\ref{sec:threatmodel}, if a local ARP spoofing or local BGP
prefix hijacking occurs, the selected verifiers will not be affected and will
thus attempt to verify the location of a machine that is different from the
fraudulent one communicating with the client.

Leaking client credentials (column 3) and TLS stripping (column 5) require
the adversary to conduct traffic hijacking first, and SLV provides partial
protection if that hijacking was locally conducted (column 2). Thus, SLV
offers a partial benefit (\prt) in both situations. The \textit{Affirms
POST-to-HTTPS} benefit prevents submitting POST requests over HTTP; SLV does
not provide that benefit.

In terms of the impact to HTTPS, no new authentication tokens are introduced
by SLV since the verification results are sent to requesting clients
automatically and in realtime. For deployability, SLV with assertions
based on IP-address to location lookup tables requires no server-side
changes, and can be deployed without DNSSEC. It provides the benefit of
\textit{Internet Scalable} because the location verification process is
fully automatable (unlike, \eg certificate preloading, where requests are
manually reviewed~\cite{kranch2015upgrading}). However, the benefit is
graded as only partial because the required verification infrastructure,
such as the verifiers, grows as the need for location verification
increases. Finally, SLV provides the benefit of signalling the status
completely because all browsed servers' locations are sent to the \manager\
(see Section~\ref{sec:methodology}) for verification, \ie server participation
is not optional.

Note that the nature of communication between a browser and the
SLV \manager\ is similar to that of the SSL multipath probing
primitive~\cite{wendlandt2008perspectives}, and thus their beneficial
properties (row 6) are similar (\cf~\cite{clark2013sok}).

{\bf Location Pinning Alternatives.}
Server location pinning is conceptually similar to key/certificate pinning,
and thus any of the four methodologies (see rows 2-5 in Table~\ref{tab:prims})
could be adopted for the server's location. For example, just as public
keys could be pinned with DNS records, servers' geographic locations can
be likewise. In fact, the DNSLOC records were proposed experimentally in
1996~\cite{rfc1876} for non-adversarial location assertion purposes. In
conclusion, evaluation outcomes of location pinning primitives are similar
to those of key pinning (\cf~\cite{kranch2015upgrading}).

Note that the benefits \textit{No New Entity}, \textit{No New Traceability},
\textit{No New Authentication Tokens}, and \textit{No Extra Third Party}
are provided by (some of) the location pinning primitives, but not SLV (row
1). For example, for \textit{No New Traceability}, the pinning process itself,
including checking verification results against already pinned locations, does
not introduce new traceability (\eg if the location verification results were
already cached). An analogous argument applies to the remaining three benefits.

{\bf List of Expected Locations.}
Domain owners can maintain a publicly accessible list of geographic
locations where a client should expect the server offering
their content. This is analogous to maintaining a \textit{list
of active certificates}~\cite{clark2013sok} (\eg Certificate
Transparency~\cite{laurie2014cert}) to facilitate revocation simply by removing
the revoked certificate from the list, and may thus aid in location revocation.

\section{Discussion}
\label{sec:discussion}

When SSL/TLS is not available for a domain, \eg if not supported or because
of an SSL stripping attack, the location verification mechanism presented
herein offers an important means (independent of SSL/TLS) for detecting
fraudulent server authentication. Nonetheless, verified location information
is best combined with SSL/TLS, to provide an additional authentication
dimension. This becomes especially useful in cases where certificate
validation is suspicious, \eg when the browser is presented a self-signed
certificate, an insecure/outdated cipher suite, or an HTTPS page with mixed
content~\cite{liang2014https}.

\begin{figure}
\centering
\includegraphics[width=0.45\textwidth]{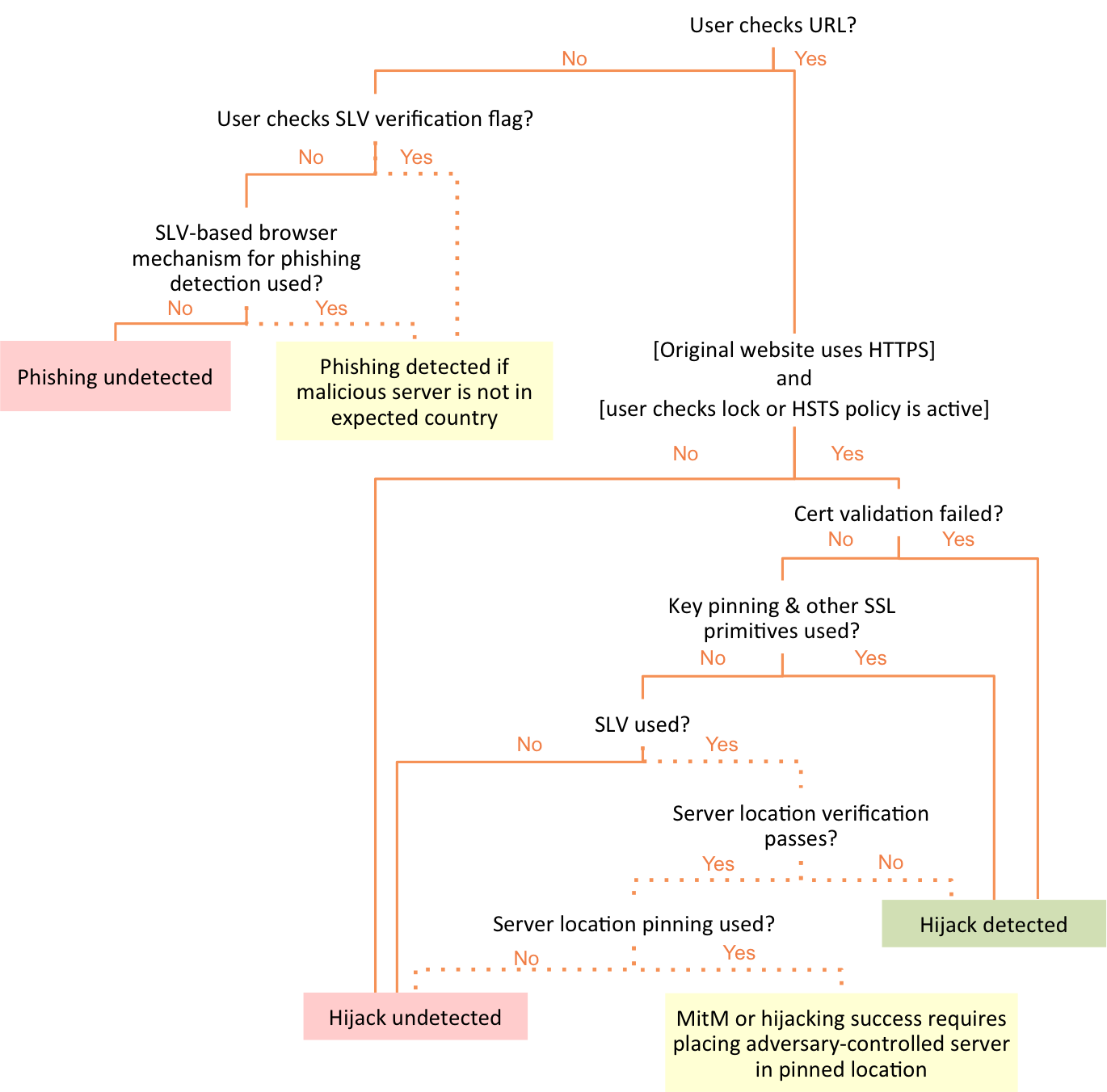}\caption{Decision
tree for detection of traffic hijacking attacks. As explained in
Section~\ref{sec:threatmodel}, from the server's perspective, phishing is a
class of traffic hijacking. Dashed lines indicate attacks detected only by
the new mechanisms presented herein.}
\label{fig:tree}
\end{figure}

The tree diagram in Fig.~\ref{fig:tree} shows how location-based server
authentication can complement SSL to reduce the likelihood of successful
attacks. The dashed lines indicate parts contributed by the presented
primitives; they highlight scenarios where traffic hijacking and/or MitM
may go undetected without SLV, but would instead be mitigated if SLV is used.

From a user's perspective, we believe that, even without requiring any new
user actions, it can be useful for some users to see in which country a server
is located---whether this information is verified by SLV or just asserted
by any browser plugin, \eg flagfox (see Section~\ref{sec:relatedwork}). In a
phishing attack, if the adversary obtains a valid certificate for a spoofed
domain, standard visual browser cues will show green locks, and positively
assuring symbols~\cite{adelsbach2005visual}. A country's flag or a displayed
world map will however differ from expectations (\ie when the adversary's
fraudulent machine is hosted remotely). There may be higher potential to
attract user notice when such a location indicator conveys intuitively
meaningful information (\eg the country flag or city where the server is)
rather than cryptic symbols---a green lock, or an exclamation mark on a grayed
out triangle, \etc. This case is similar to that where a browser-trusted CA
is compromised and no certificates are pinned for the victim domain. On the
other hand, location verification provides that missing benefit of signaling
such an adversarial situation using an intuitively meaningful visual cue; \eg
the browser will either display an unexpected flag if the asserted country is
different from that of the authentic server, or depending on implementation
choices, show a struck out flag in case the location fails verification.

Note that several useful features can be built on top of server
location verification. For instance, SLV can benefit from a policy-based
mechanism~\cite{wendlandt2008perspectives,santos2012policy} that customizes
how a browser automatically handles various transactions based on their
location~\cite{evilCities}. An instruction could be of the form \emph{allow
credit card transactions only at this set of countries} or \emph{deny email
logins in that set of locations}. This may also help control fraud and
deter phishing attacks, ideally requiring no new user actions or decisions
whatsoever. Also, any of several known mechanisms could be employed to help the
system's scalability and efficiency (\cf \cite{bates2014forced}). For example,
the verifiers (acting as regular clients) could proactively measure delays
periodically (e.g., to high-runner websites) to reduce verification time.

\section{Related Work}
\label{sec:relatedwork}

{\bf GeoPKI.} GeoPKI \cite{kim2013geopki} is a location-aware PKI system
that associates certificates to geographic \emph{spaces}, \eg land or
property boundaries. A certificate contains a high granularity definition
of the space to which it is associated. This could be in the form of GPS
coordinates along with lateral and longitudinal distances that accurately
delineate the space boundaries. To claim a space, the owner submits their
space-defined certificate (self-signed or CA-signed) to a public log (\eg
similar to certificate transparency \cite{laurie2014cert}), and monitors
the log to detect any other entity claiming ownership of their space. To
validate a space ownership, GeoPKI relies on CA-issued Extended Validation
(EV) certificates, associated to a real world street address. An attacker
would thus need to either compromise a CA to issue an EV certificate to tie
its public key to the fraudulently-claimed space, or forge legal documents
proving such ownership.

The goals and threat model of GeoPKI differ from those we address
herein. GeoPKI provides no indication or assurance of the location of the
actual server a client is connected to.\footnote{A client could be connected
to a webserver in China, legally owned and operated by an entity in the
US. The GeoPKI EV certificate then validates the US location, with no user
awareness of the physical server's location.} Additionally, no realtime
location verification mechanism is involved, thus a compromised CA remains
a threat. With SLV, compromising a CA alone is insufficient; the attacker
must also evade SLV to succeed in a MitM attack.

{\bf Client Presence Verification.} CPV~\cite{cpvtdsc} verifies geographic
location claims of Internet \emph{clients}. Client locations are corroborated
based on triangular \emph{areas} derived from delay measurements. While
this enables CPV to verify client locations with high granularity, its
verification process suffers occasional Triangular Inequality Violations
(TIVs)~\cite{tivrecent}. In contrast, while SLV selects verifiers forming
triangles, it does not verify server presence within them, nor use triangular
areas; its use of Thales's theorem avoids TIVs entirely, and reduces the
number of delay measurements required between verifiers and the webserver. SLV
also provides assurance to clients about \emph{server} locations (note that
servers and clients differ fundamentally in many factors, including that a
third party location verification service provider can easily get clients
to run code, \eg using JavaScript as CPV does; this is not applicable when
verifying geographic locations of webservers).

{\bf Geolocating fast-flux servers.} Delay-based geolocation of fast-flux
hidden webservers has been proposed~\cite{proxiedservices}; hidden behind
proxies, their IP addresses are not known to the client. When geolocating
a webserver, the geolocation service provider can first detect that the
webserver is hidden behind a proxy by noticing a large difference between
the RTTs measured on the network layer (\eg using \texttt{ping}) and the
application layer (\eg using an HTTP GET). To estimate the hidden server's
location, a group of landmarks measure application layer RTTs to the server,
which are then used to obtain rough estimates to the direct RTTs between the
landmarks and the hidden server (excluding the proxy). The RTTs are then
mapped to distances to constrain the region of the hidden server relative
to the landmarks \cite{Constrainbased}.

Such a geolocation mechanism aims at disclosing an inconsistency between
the geographic location of the sever terminating the TCP connection and the
one processing HTTP requests. SLV does not attempt to determine webserver
locations, but rather verifies the plausibility of the webserver within an
asserted region. While attempts to evade SLV may include hiding the attacker's
IP address behind a proxy, SLV handles that attack differently---it reports
that the asserted location (that of the IP address seen by the client)
is not verified.

{\bf Flagfox extension.} Flagfox is an example\footnote{Other extensions exist
with similar objectives.} Firefox extension that looks up the countries of
webserver IP addresses as a user browses the Internet, and displays the country
flag in the URL bar. The flag is based on the tabulated location of the IP
address, not the country TLD in the domain name. Flagfox uses Maxmind's IP
database for geolocation,\footnote{\url{https://www.maxmind.com}} and does
not employ any location verification mechanism. Since locations obtained
by tabulation-based techniques are falsifiable~\cite{MuirPaul}, \eg by
the IP address owner, they are unreliable in adversarial environments. For
instance, an adversary aiming to impersonate the University of Tennessee's
website (\eg through phishing, pharming or a MitM attack) could register
the IP address assigned to its malicious webserver to be in Knoxville,
Tennessee. Indeed, a previous study~\cite{laki2011spotter} found that most
of Google's IP addresses are reported by the American Registry for Internet
Numbers (ARIN)\footnote{\url{http://whois.arin.net}} to be physically located
in Mountain View, California; such clearly incorrect assertions have been
proven wrong~\cite{laki2011spotter}.

\section{Concluding Remarks}
\label{sec:conclusion}

The server location verification mechanism detailed herein does not
conflict with the web's growing trend of distributed content dissemination
and geographically diverse replicated caching. The initial front-end
server to which a client connects is the port of entry to the distribution
infrastructure, if one is being used; paying more attention to that server,
\eg by verifying its physical presence in a known/expected geographic location
as explained herein, provides information often relevant to the target domain's
authenticity. Thus, SLV works regardless of the distribution and architecture
of such infrastructure. Additionally, depending on a client's location, a
finite set of $n$ such ``ports of entry" are typically expected for any single
domain, and that set is often stable. Pinning several server locations (see
Section~\ref{sec:pinning}) is thus beneficial for new and verified locations.

Despite efforts from the security community to address shortcomings in
the SSL/TLS ecosystem, PKI compromise still admits MitM attacks, \eg due
to slow or non-adoption of primitives like key pinning, or user inability
to reliably react to visual browser cues. The proposals herein constitute
a new and parallel server authentication dimension augmenting SSL/TLS (\eg
comparable to client multi-factor authentication), relying not on the standard
\emph{something you have} principle (namely, the server private key), but
in addition \emph{where you are}. While the general notion of location-based
authentication is known, the novelty herein is the measurement-based mechanism
itself which verifies server locations in realtime, in a manner compatible
with SSL/TLS authentication, and without requiring human-user involvement
in decision making.

To mount a successful MitM attack when SLV is used, the adversary must, in
addition to compromising the SSL/TLS infrastructure, co-locate its malicious
(possibly virtual) machine in the geographic vicinity of the authentic
one. This places a heavy set of burdens, including an attack customized to
the location of each target server. A mechanism like SLV thus compels the
adversary to make a true assertion about the location of its fraudulent
servers, both divulging the fraudulent servers' true geographic location,
and forcing the adversary to operate in the geographic vicinity of the
authentic webserver---often a region in a more familiar country, or with
more favourable laws and accountability measures.

SLV leverages established networking principles that location information can
be inferred from timing measurements, and existing methodological guidelines
for use of timing measurements to achieve server location verification. While
large-scale evaluation of SLV's verification process is not the main focus
of this paper, preliminary experiments highlight the algorithm's efficacy in
verifying webservers' geographic locations, by means immediately deployable
through a browser extension without requiring webserver modifications.

\ifCLASSOPTIONcompsoc
  \section*{Acknowledgments}
\else
  \section*{Acknowledgment}
\fi
The second author acknowledges funding from the Natural Sciences and
Engineering Research Council of Canada (NSERC) for both his Canada Research
Chair in Authentication and Computer Security, and a Discovery Grant.

\end{document}